\documentclass[twocolumn,amssymb, amsmath, showpacs, superscriptaddress, aps, prl]{revtex4}
\usepackage{latexsym}        
\usepackage{amssymb}
\usepackage{amsmath}
\usepackage{amsfonts}
\usepackage{graphicx} 
\usepackage{color}

\begin{document}
\title{Temperature dependence of iron local magnetic moment in phase-separated superconducting chalcogenide}

\author{L. Simonelli}
\affiliation{European Synchrotron Radiation Facility, BP220, F-38043 Grenoble Cedex, France}
\affiliation{CELLS - ALBA Synchrotron Radiation Facility, Carretera BP 1413, Km 3.3
08290 Cerdanyola del Valles, Barcelona, Spain}

\author{T. Mizokawa}
\affiliation{Dipartimento di Fisica, Universit{\'a} di Roma ``La Sapienza" - P. le Aldo Moro 2, 00185 Roma, Italy}
\affiliation{Department of Complexity Science and Engineering \& Department of Physics, University of Tokyo, 5-1-5
Kashiwanoha, Kashiwa, Chiba 277-8561, Japan}

\author{M. Moretti Sala}
\affiliation{European Synchrotron Radiation Facility, BP220, F-38043 Grenoble Cedex, France}

\author{H. Takeya}
\affiliation{National Institute for Materials Science, 1-2-1 Sengen, Tsukuba 305-0047, Japan}

\author{Y. Mizuguchi}
\affiliation{Dipartimento di Fisica, Universit{\'a} di Roma ``La Sapienza" - P. le Aldo Moro 2, 00185 Roma, Italy}
\affiliation{National Institute for Materials Science, 1-2-1 Sengen, Tsukuba 305-0047, Japan}
\affiliation{Department of Electrical and Electronic Engineering, Tokyo Metropolitan University, 1-1 Minami-osawa, Hachioji, Tokyo, 192-0397, Japan}

\author{Y. Takano}
\affiliation{National Institute for Materials Science, 1-2-1 Sengen, Tsukuba 305-0047, Japan}

\author{G. Garbarino}
\affiliation{European Synchrotron Radiation Facility, BP220, F-38043 Grenoble Cedex, France}

\author{G. Monaco}
\affiliation{European Synchrotron Radiation Facility, BP220, F-38043 Grenoble Cedex, France}
\affiliation{Dipartimento di Fisica, Universit{\'a}́ di Trento, I-38123 Povo, Trento, Italy}

\author{N.L. Saini}
\affiliation{Dipartimento di Fisica, Universit{\'a} di Roma ``La Sapienza" - P. le Aldo Moro 2, 00185 Roma, Italy}

\date{\today}

\begin{abstract}
We have studied local magnetic moment and electronic phase separation in superconducting K$_{x}$Fe$_{2-y}$Se$_2$ by x-ray emission and absorption spectroscopy. Detailed temperature dependent measurements at the Fe K-edge have revealed coexisting electronic phases and their correlation with the transport properties. By cooling down, the local magnetic moment of Fe shows a sharp drop across the superconducting transition temperature (T$_c$) and the coexisting phases exchange spectral weights with the low spin state gaining intensity at the expense of the higher spin state. After annealing the sample across the iron-vacancy order temperature, the system does not recover the initial state and the spectral weight anomaly at T$_c$ as well as superconductivity disappear. The results clearly underline that the coexistence of the low spin and high spin phases and the transitions between them provide unusual magnetic fluctuations and have a fundamental 
role in the superconducting mechanism of electronically inhomogeneous K$_{x}$Fe$_{2-y}$Se$_2$ system.
\end{abstract}


\maketitle
\section{Introduction}
One of the interesting discoveries in the iron-based chalcogenides has been a successful intercalation of FeSe by alkaline atoms. The intercalated FeSe, forming a new A$_{x}$Fe$_{2-y}$Se$_2$ (A = K, Rb, Cs) family, shows a T$_c$ as high as 32 K ~\cite{Guo,Mizuguchi,Ying,Ming-Hu} and, more importantly, exhibits several unique microstructural characteristics, such as a large magnetic moment per Fe site, intrinsic Fe vacancy order in the $ab$-plane and an antiferromagnetic order in the $c$-direction~\cite{Ryan,Bao,Bao2}. In addition, A$_{x}$Fe$_{2-y}$Se$_2$ also shows phase separation ~\cite{Ricci,Wang2,Chen}, in which a magnetically ordered phase with an expanded lattice and a nonmagnetic phase with a compressed lattice coexist ~\cite{Ricci}. The phase separation in A$_{x}$Fe$_{2-y}$Se$_2$ has been studied by several techniques ~\cite{Bao2,Ricci,Wang2,Wang,Guo2,Chen,Zhang,Yan,Han,TexiNMR,SherMuSR,HomeOpt,YuanOpt,KsenMoss,Shoemaker12,Bendele14,Oiwake13,Simonelli12},  probing different properties. In particular, detailed temperature dependent 
neutron diffraction studies on superconducting A$_{x}$Fe$_{2-y}$Se$_2$ (A=K, Rb, Cs, Tl) have revealed same Fe vacancy ordered crystal structure and the same block antiferromagnetic order in all these materials ~\cite{Bao2}. The intensity of the magnetic Bragg peak, providing information on the iron magnetic moment, shows a continuous increase by lowering temperature, however, this behavior is interrupted by a flat plateau when T$_c$ is approached. Muon spin relaxation ($\mu$SR) experiments have shown phase separation with $\sim$88-90\% of the sample volume to exhibit large magnetic moment while $\sim$10-12\% of the sample to remain paramagnetic \cite{SherMuSR}. Similar conclusions have been reached by  Nuclear magnetic resonance (NMR) experiments \cite{TexiNMR} showing distinct spectra originating from a majority antiferromagnetic and a minority metallic phase. The NMR results have also revealed a sharp drop in the relaxation rate and Knight shift on cooling across T$_c$, indicating development of a field distribution due to the appearance of a vortex state. Optical properties of superconducting K$_{x}$Fe$_{2-y}$Se$_{2}$, have shown that $\sim$10\% sample volume is covered by highly distorted metallic inclusions in an insulating matrix, suggesting a filamentary network of conducting regions \cite{HomeOpt} and Josephson-coupling plasmons \cite{YuanOpt}. In addition, $^{57}$Fe-M$\ddot{o}$ssbauer spectroscopy has revealed phase separarion in $\sim$88\% magnetic and $\sim$12\% nonmagnetic Fe$^{2+}$ species \cite{KsenMoss}. A recent space resolved photoemission has provided further information on the peculiar electronic phase separation in metallic filamentary phase and majority insulating texture \cite{Bendele14}. Indeed, It is now known that the majority phase is insulating with a stoichimetry of  A$_{0.8}$Fe$_{1.6}$Se$_2$ (245) containing iron vacancy order and block antiferromagnetism with large magnetic moment ($\sim$3.3 $\mu_B$) while the minority phase is metallic with a stoichiometry of A$_{x}$Fe$_{2}$Se$_2$ (122). 

Very recently, the coexisting electronic phases in K$_{x}$Fe$_{2-y}$Se$_2$ have been studied using high energy spectroscopy ~\cite{Simonelli12}, revealing different phases to be characterized by different Fe valence and local magnetic moment. 
In this work, we have further exploited x-ray emission (XES) and high resolution x-ray absorption (XAS) spectroscopy to study correlation between the superconductivity and the electronic/magnetic phase redistribution as a function of temperature.
While XES has been used to investigate the evolution of local Fe magnetic moment $\mu$, XAS has been exploited to study unoccupied electronic states, corresponding to different electronic phases. The magnetic moment $\mu$ shows anomalous temperature dependence with a sharp drop at the superconducting transition temperature. The results also reveal
that the coexisting electronic phases with Fe$^{2+}$ high spin (HS) and low spin (LS) exchange their spectral weights across the superconducting transition temperature ($\sim$32 K). On the other hand, by annealing the sample across the phase separation and iron-vacancy ordering temperatures ($\sim$520 K and $\sim$580 K respectively), the mean local moment $\mu$ is decreased, and the HS-LS spectral exchange across T$_c$ gets  suppressed with disappearance of the superconductivity. 

\section{Experimental details}\label{S:Exp}
The XES and high resolution XAS measurements were carried out on well characterized single crystals of K$_x$Fe$_{2-y}$Se$_2$.  After the growth using the Bridgman method, the crystals were sealed into a quartz tube and annealed for 12 hours in vaccum at 600$^\circ$C. The crystals were characterized for their structure and phase purity by x-ray diffraction measurements ~\cite{Mizuguchi}. The electric and magnetic characterizations were performed by resistivity measurements in a PPMS (Quantum Design) and magnetization measurements in a SQUID magnetometer (Quantum Design). The samples exhibit a sharp superconducting transition at $T_{\rm c}\simeq 32$ K with a transition width of $\sim$1 K in the temperature dependent resistivity signal. The superconducting volume fraction was found to be $\sim$10\% at 2 K. The samples were exposed to air only for a couple of minutes while mounting  on the sample holder. For the rest of the time the samples were manipulated in a glovbox or were kept in vacuum. 

The spectroscopy measurements were performed at the beamline ID16 of the European Synchrotron Radiation Facility. The experimental setup consists of a spectrometer based on the simultaneous use of a bent analyzer Ge(620) crystal (bending radius R = 1 m) and a pixelated position-sensitive Timepix detector~\cite{Ponchut} in Rowland circle geometry.  The scattering plane was horizontal and parallel to the linear polarization vector of the incident x-rays.  The measurements were carried out fixing the sample surface~(the $ab$-plane of the K$_x$Fe$_{2-y}$Se$_2$ single crystal)~at~$\sim$~45$^{\circ}$~from the incoming beam direction and the scattering angle 2$\theta$ at $\sim$ 90$^{\circ}$. The total energy resolution was about 1.1 eV~full width at half maximum.  For the low temperature measurements the sample was placed in a cryogenic environment, while for the high temperature measurements it was placed in an oven. Firstly, we have cooled down the sample from room temperature (RT$\sim$300 K) to 18 K, across T$_c$. This was followed by warming the sample up to 600 K, destroying the phase separation (at $\sim$520 K) and disordering the Fe vacancies (at $\sim$580 K). At the end, we have cooled down the sample again to 18 K. The sample temperatures during the measurements were controlled with an accuracy of~$\pm$1~K.
 
\section{Results and discussions}\label{S:analysis}
\begin{figure}
\includegraphics[width=0.9\linewidth]{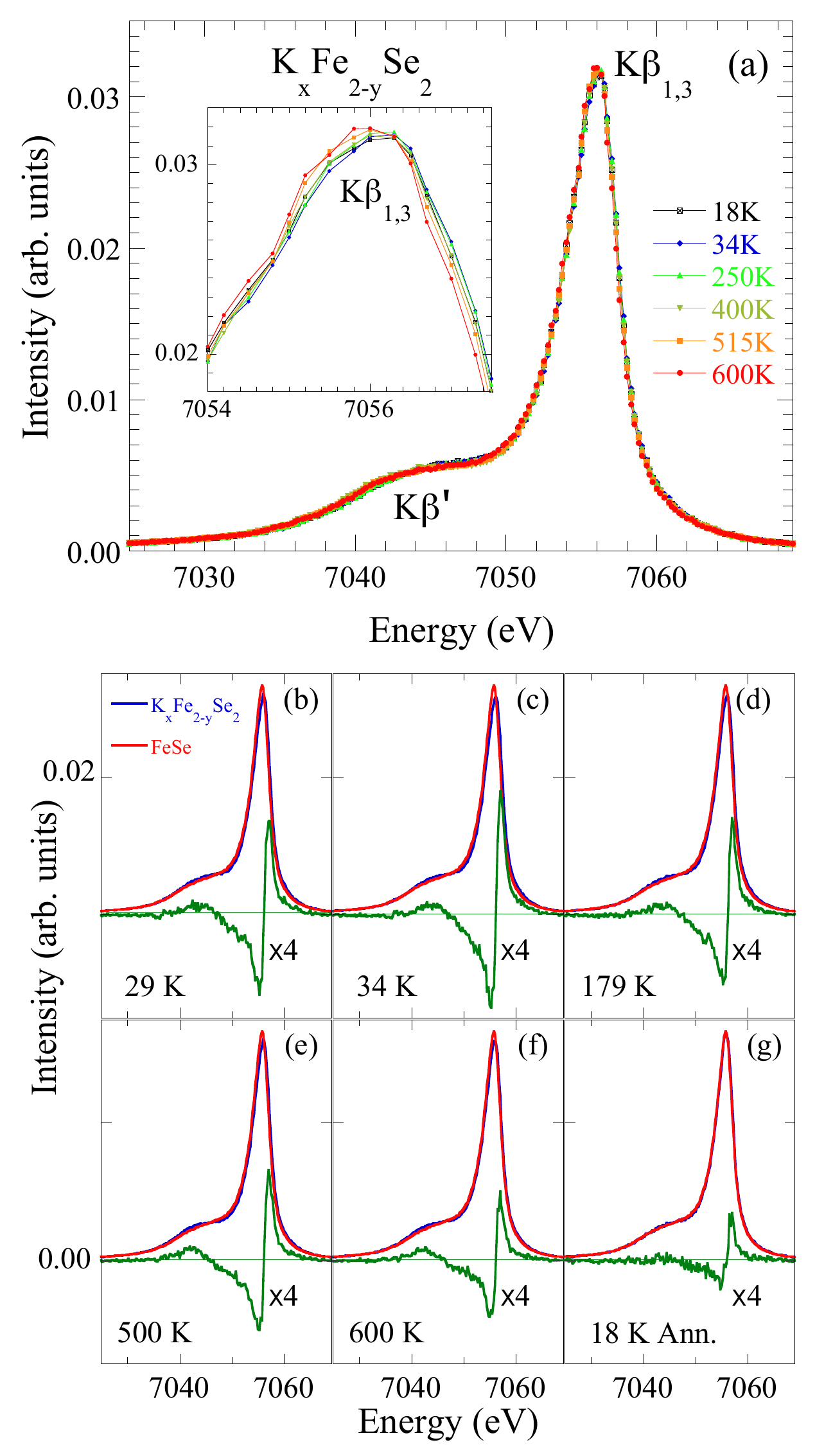}
\caption{(a) Fe K$\beta$ XES measured on K$_x$Fe$_{2-y}$Se$_2$ at several temperatures. The spectra are normalized to the integrated area of the XES lines. The inset shows a zoom over the main K$\beta_{1,3}$ emission peak. (b)-(g)  Fe K$\beta$ emission spectra of K$_x$Fe$_{2-y}$Se$_2$ (blue) at several temperatures are shown compared with the one measured on FeSe at RT (red). The difference spectra (green) are also shown (multiplied by four).}
\label{F:spectra}
\end{figure}

Figure~1~(a) shows Fe K$\beta$ emission spectra measured on K$_x$Fe$_{2-y}$Se$_2$ at different temperatures. In the crystal-field picture the overall spectral shape is dominated by the ($3p$,$3d$) exchange interactions~\cite{Glatzel}, and the presence (absence) of a pronounced feature at lower energy (K$\beta$') is an indication of a HS (LS) state of Fe~\cite{Glatzel}. The energy position of the K$\beta_{1,3}$ provides similar information on the spin state reflecting the effective number of unpaired $3d$ electrons~\cite{Glatzel}. A direct comparison between measured spectrum on FeSe with those on  K$_x$Fe$_{2-y}$Se$_2$ (see, e.g. Fig. 1(b) to 1(g)) shows that the local Fe magnetic moment $\mu$ is higher for the latter. It should be noted that the $\mu$ for the present sample is smaller than the one for an air-exposed K$_x$Fe$_{2-y}$Se$_2$ sample~\cite{Simonelli12}. This further underlines fragile nature of K$_x$Fe$_{2-y}$Se$_2$ morphology with physical properties being very sensitive to the air exposure. A zoom over the main K$\beta_{1,3}$ emission off the  K$_x$Fe$_{2-y}$Se$_2$ is shown as inset of Fig. 1(a). Small differences are apparent as a function of temperature. These variations are due to changing local Fe magnetic moment $\mu$~\cite{Glatzel,Cowan,Vanko,Gret13}. 

\begin{figure}
\includegraphics[width=0.9\linewidth]{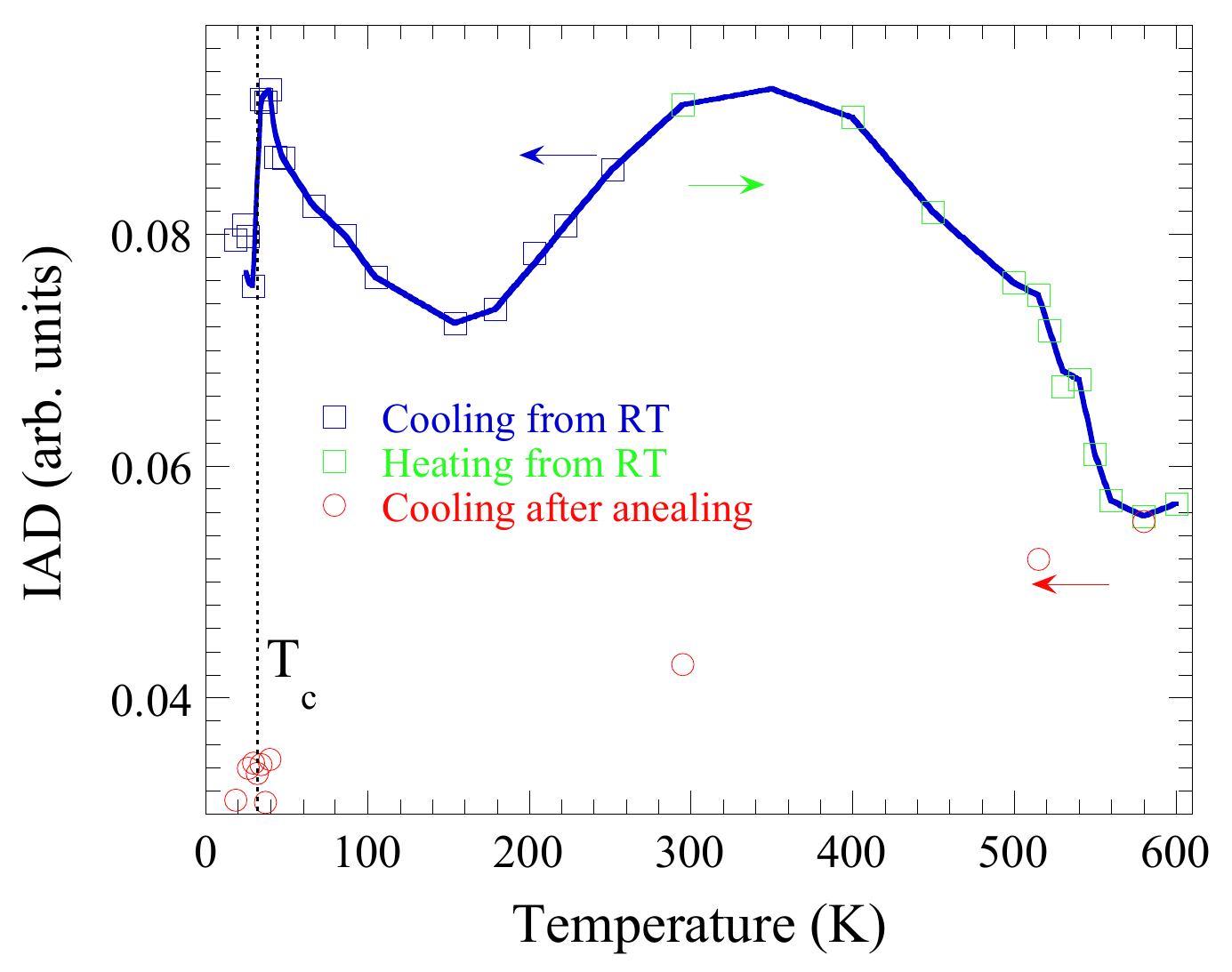}
\caption{Integrated absolute difference (IAD) calculated with respect to the FeSe as a function of temperature (see text). The thick continuous line (blue) is a guide to the eyes. The size of the symbols represents maximum uncertainty, estimated by analyzing different XES scans at each temperature.}
\label{F:spectra}
\end{figure}

Integrated absolute difference (IAD) of the K$\beta$ spectra with respect to a non magnetic reference is proportional to the local $\mu$~\cite{Gret13}. Here, we have determined IAD relative to the spectrum of FeSe.  The FeSe is not a non-magnetic reference, however, the IAD still provides useful information on the relative evolution of $\mu$ as a function of temperature. Figure 1 (b)-1(g) shows Fe K$\beta$ emission spectra of K$_x$Fe$_{2-y}$Se$_2$ at different temperatures compared with the one measured on FeSe. The difference spectra provides information of the changing local moment. The difference spectra evolve with temperature showing some apparent changes across T$_{c}$ [Fig. 1(b) and 1(c)].  A substantial change can be seen [Fig. 1(e) and 1(f)] across the temperature for iron vacancy order-disorder ($\sim$580 K). These variations are suppressed while the sample is cooled after the annealing above the iron vacancy disorder temperature [Fig. 1(g)].

Detailed temperature evolution of the IAD is shown in Figure 2. By cooling from RT, the IAD shows an anomalous behavior, representing an abnormal evolution of $\mu$. Indeed, the IAD shows a gradual decrease and an increase (with a local cusp-like anomaly at $\sim$ 170 K) before showing a sharp drop at T$_c$. This is a clear indication of a lowering of the local magnetic moment in the superconducting state. 
Earlier neutron diffraction studies on superconducting A$_{x}$Fe$_{2-y}$Se$_2$ have shown that the temperature dependent magnetic Bragg peak intensity exhibits a plateau below T$_c$ \cite{Bao2}, indicating 
an interesting interplay between the ordered magnetic moment and superconductivity. In the present work, the local magnetic moment probed by the XES is found to show a sharp drop at T$_c$, revealing that there is HS-LS transition and the resulting magnetic fluctuations have a direct relation with the superconductivity in phase separated K$_{x}$Fe$_{2-y}$Se$_2$.

The IAD shows a gradual decrease with increasing temperature, revealing a transition-like behavior around the phase separation and iron-vacancy disorder temperatures ($\sim$520 K and 580 K respectively). Indeed, by increasing temperature from RT to 600 K the K$\beta_{1,3}$ emission line shifts towards lower energy [Fig. 1(a)] and the K$\beta'$ spectral weight is suppressed. These results suggest that the system is undergoing a change in the spin configuration across the phase separation temperature of $\sim$520 K~\cite{Ricci}, reaching a minimum around 580 K [Fig. 2] where the block antiferromagnetic order is known to get suppressed. Cooling down to RT, the sample does not recover the initial conditions, consistent with a glassy nature \cite{Iadecola12}. By cooling after the annealing, the discontinuity across T$_{c}$, evident in the cooling cycle, does not appear. Magnetic susceptibility measurements on the annealed sample does not reveal any significant superconducting signal. This indicates that, either the annealed sample is no more superconducting or the superconducting volume fraction is significantly reduced.

\begin{figure}
\includegraphics[width=0.9\linewidth]{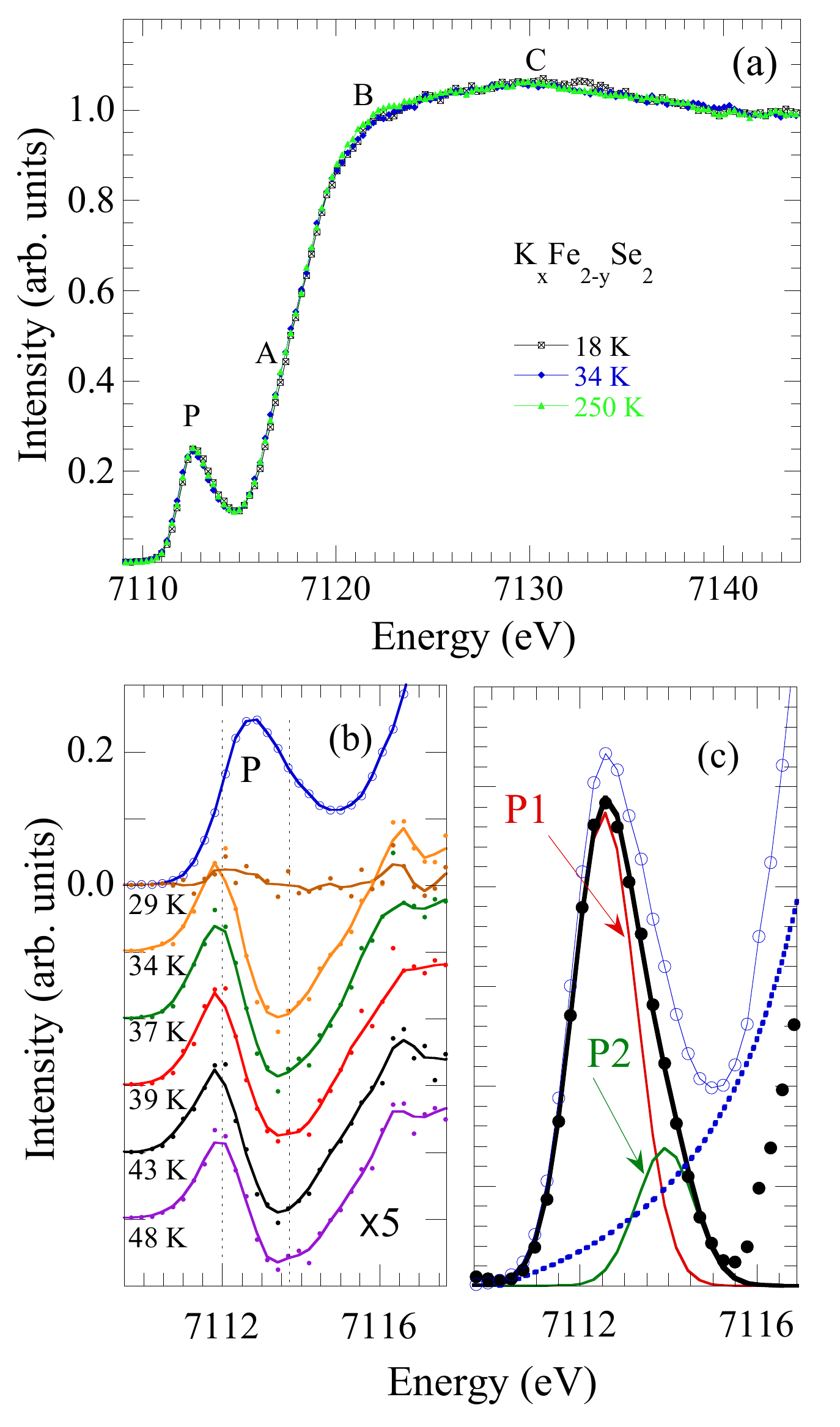}
\caption{(a) Partial fluorescence yield (PFY) XAS spectra measured at the Fe K-edge of K$_x$Fe$_{2-y}$Se$_2$ at various temperatures. (b) A zoom over the pre-peak region with spectral differences across the transition temperature (multiplied by five) with respect to a spectrum measured below the transition. (c) Deconvoluted absorption pre-edge peak in two components (P1 and P2) after subtracting a constant arctangent background from the normalized absorption (the latter are also shown).}
\label{F:spectra}
\end{figure}

An independent and complementary information can be obtained from XAS. Figure 3 shows the Fe K-edge partial fluorescence yield (PFY) XAS spectra measured on K$_x$Fe$_{2-y}$Se$_2$ at different temperatures. The spectra are measured by collecting the Fe K$\beta_{1,3}$ emission as a function of incident energy, and normalized with respect to the atomic absorption estimated by a linear fit far away from the absorption edge. The XAS spectra show an asymmetric single pre-edge peak P~\cite{Simonelli12,Simonelli,Joseph}, a shoulder A and broad near edge features B and C. The broad features indicate the system to have large atomic disorder, consistent with local structural studies ~\cite{Iadecola12}. The main peak B depends on temperature indicating a change in the Fe 4$p$-Se 5$d$ hybridization in K$_x$Fe$_{2-y}$Se$_2$ \cite{Simonelli,Joseph}. On the other hand, pre-edge peak exhibits small temperature dependence, indicating change in the unoccupied Fe 3$d$-Se 4$p$ electronic states~\cite{Simonelli,Joseph} as a function of temperature. 

\begin{figure}
\includegraphics[width=0.9\linewidth]{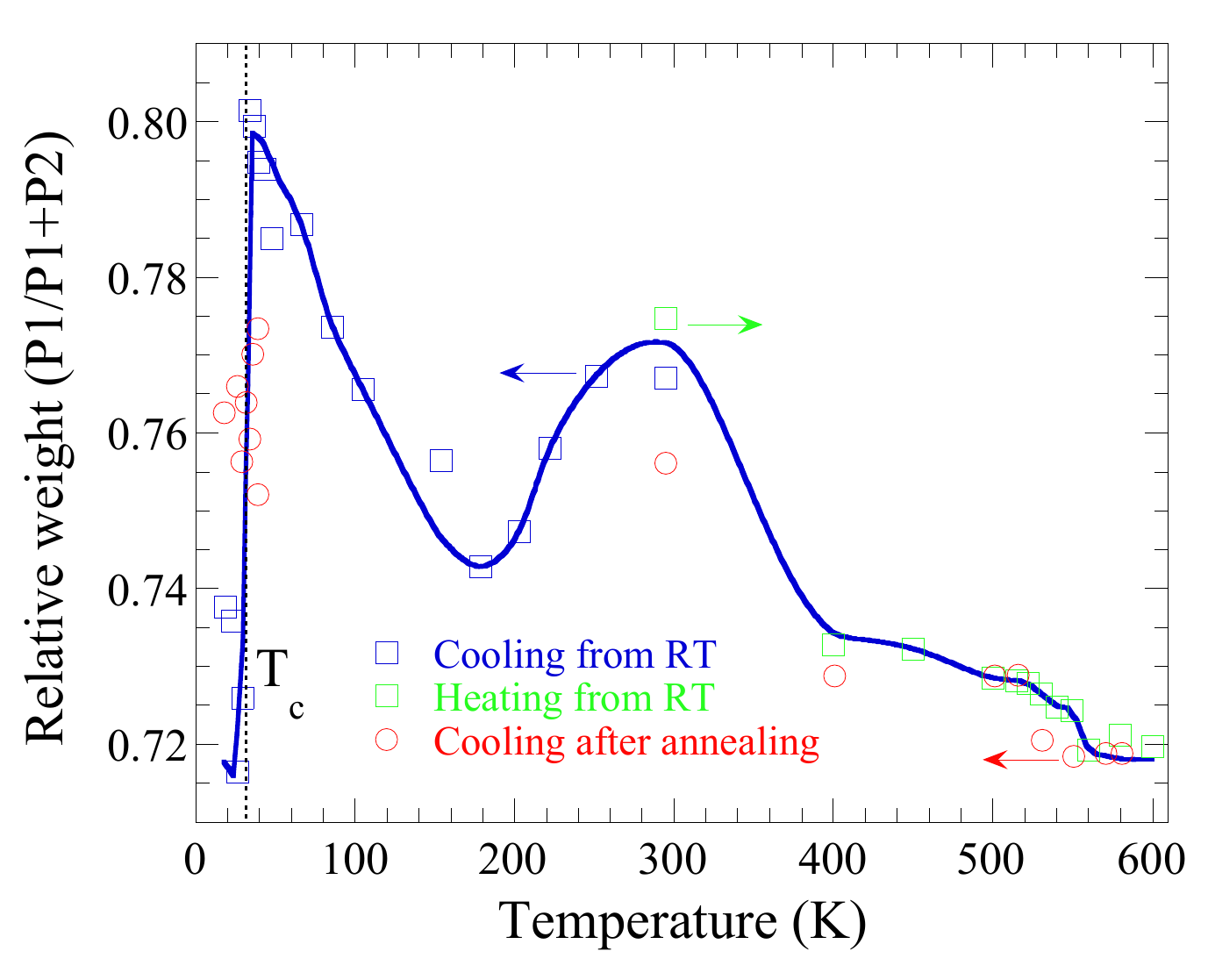}
\caption{Relative weight of the P1 [Fig. 3(c)] as a function of temperature. 
The thick continuous line (blue) is a guide to the eyes. The size of the symbols represents maximum uncertainty, estimated by Guassian fits of different spectra at each temperature.}
\label{F:spectra}
\end{figure}

In order to describe the temperature evolution of the electronic states near the Fermi level, Fig. 3 (b) shows difference spectra across T$_c\sim$32 K with respect to a spectrum measured (at 23 K) well below T$_c$. The difference spectra reveal spectral weight redistribution from the lower energy side to the higher energy side of the pre-edge peak. To evaluate the temperature dependent redistribution of the spectral weight, we have performed a curve fitting of the background subtracted XAS spectra with two Gaussian functions in the pre-edge peak region. The background is given by a constant arctangent function approximating the atomic absorption. Two Gaussian functions of constant width and energy position were enough to describe the background subtracted pre-edge peak P. We have shown the zoomed absorption spectrum in the pre-edge peak region containing the background function along with the two Gaussian components (P1 and P2) of the background subtracted spectrum and the fit as Fig. 3(c). 

Let us discuss the possible origin of the two Gaussian peaks P1 and P2. It is known that Fe K-edge absorption pre-peak is sensitive to the Fe spin state in iron complexes ~\cite{HSeLS}. In addition, earlier high pressure XAS experiments on FeSe and Rb$_x$Fe$_{2-y}$Se$_2$ have revealed a crossover from HS Fe$^{2+}$ to LS Fe$^{2+}$, evidenced by an appearance of a LS Fe$^{2+}$ peak structure at $\sim$ 1 eV higher in energy than the HS Fe$^{2+}$ peak ~\cite{Chen11,Kumar11,Bendele13a}. We are also aware of the fact that different phases coexist in the superconducting K$_x$Fe$_{2-y}$Se$_2$: (i) a majority semiconducting and magnetic 245-phase with iron vacancy order, and; (ii) a metallic and nonmagnetic disordered 122- phase. The coexisting phases are characterized by slightly different Fe oxidation states, i.e., 2+, for the 245, and slightly less than 2+, for the 122 phase~\cite{Shoemaker12}. Small differences in the oxidation state are hardly detectable by the main rising absorption edge that appears to be influenced by the glassy local structure~\cite{Iadecola12}. The 245 and the 122 phases have similar structure with the latter being slightly compressed in the plane and expanded in the out of plane ~\cite{Ricci}. These phases also have different iron spin state ~\cite{Simonelli12}, with the majority phase being in HS Fe$^{2+}$ state while the minority phase appears to be in LS state. Therefore, we can assign qualitatively the two Gaussian components P1 and P2 to HS Fe$^{2+}$ majority phase and LS Fe$^{2+}$ minority phase. Considering the fact that the minority and majority phases are known to contribute about 10-20\%  and 80-90\% respectively, the spectral weight of the majority phase (minority phase) with HS (LS) Fe$^{2+}$ state counting about 78\% (22\%) at RT is in fair agreement with the present assignment. 

Temperature evolution of the relative spectral weight of the majority phase (defined as the ratio of the intensity of P1 to the total intensity, i.e., P1+P2) is shown in Fig. 4. The HS phase tends to decrease with decreasing temperature from 300 K to 180 K, revealing a cusp-like anomaly around 180 K with maximum weight being $\sim$80$\%$ before a sharp drop to $\sim$70\% around the T$_c$. On the other hand, with increasing temperature from 300 K the spectral weight of the HS phase gradually decreases with a small discontinuity above the phase separation temperature of $\sim$520 K and iron vacancy disorder temperature of $\sim$580 K. In general, temperature evolution of the relative weight of the majority phase appears consistent with the IAD representing variation of the local Fe moment $\mu$ [Fig. 1]. This justifies the assignment of the P1 and P2  due to the two coexisting phases, characterized by HS and LS Fe$^{2+}$ states, with HS phase contributing about 70-80$\%$ of the total while the remaining phase should be due to LS Fe$^{2+}$ phase. This affirmation agrees with different experimental studies showing that the magnetic phase is the majority phase while the minority phase should be one that becomes superconducting by cooling~\cite{Ming-Hu,Yan,Han}. Again, it is interesting to note that, by cooling from 600 K to 400 K, the system does not recover the initial state. Indeed, after the annealing, the discontinuity across T$_{c}$ disappears, likely to be due to suppression of superconductivity or highly reduced volume fraction of the superconducting phase. 

\begin{figure}
\includegraphics[width=\linewidth]{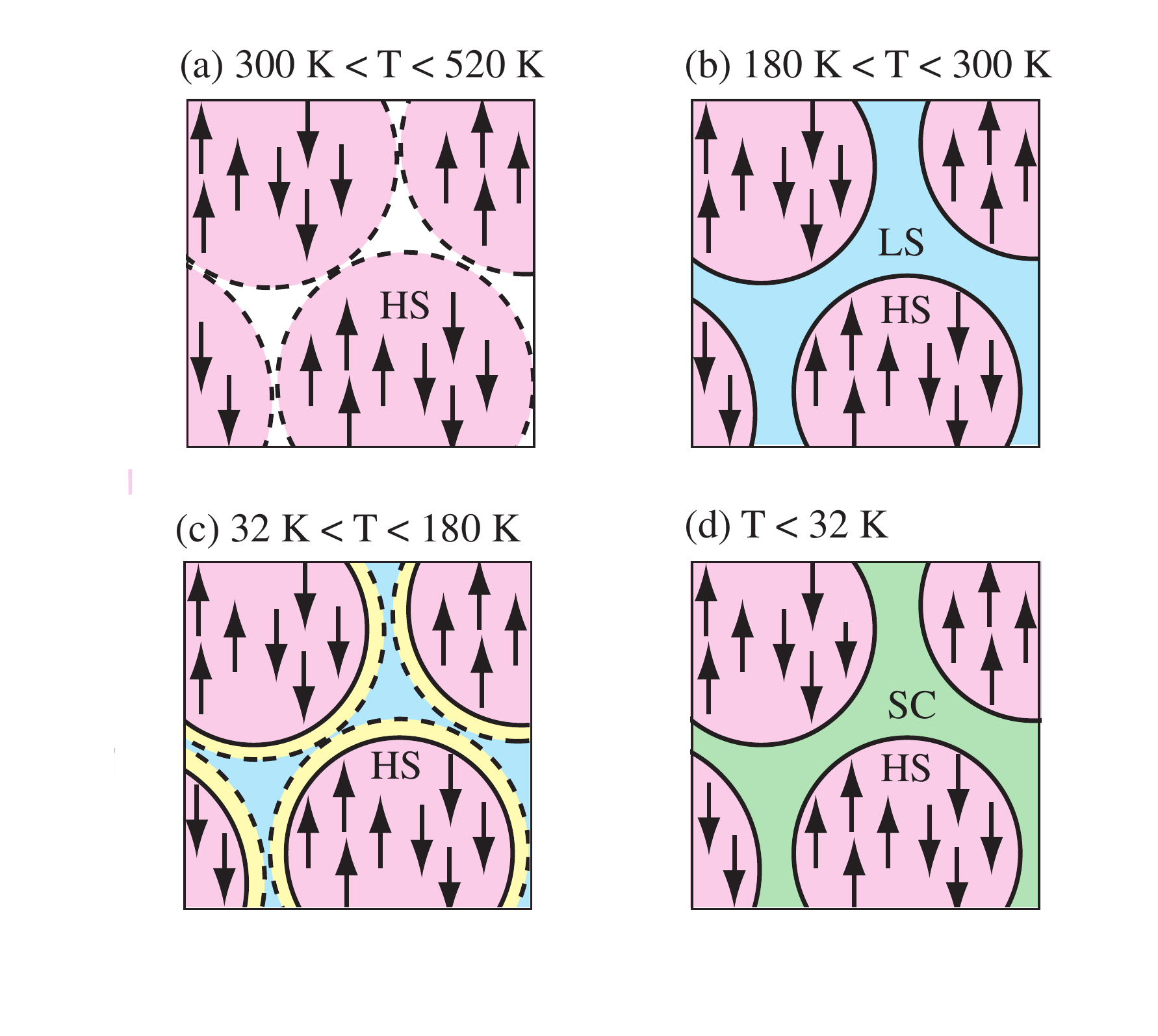}
\caption{
(a) Fe$^{2+}$ HS domains gradually grow with cooling from 520 K to 300 K. (b) Filamentary Fe$^{2+}$ LS domain is established with cooling from 300 K to 180 K. (c) The interface region between the Fe$^{2+}$ HS and Fe$^{2+}$ LS domains becomes magnetic with cooling from 180 K to 32 K. (d) The magnetic interface state turns into the superconducting state with Fe$^{2+}$ LS.}
\label{}
\end{figure}

Let us attempt to understand the XES and XAS findings shown in Fig. 2 and Fig. 4, revealing an abnormal evolution of local Fe magnetic moment and different electronic phases. We start our discussion for the cooling cycle from 300 K. It is clear that the local moment shows a decrease [Fig. 2], consistent with the decrease of the HS Fe$^{2+}$ phase [Fig. 4]. With further cooling, the Fe magnetic moment shows an increase before showing a sharp drop at the superconducting transition. This is again consistent with the evolution of the HS Fe$^{2+}$ phase fraction. However, the question is what is the origin of such an anomalous behavior of the phase fractions. 

Below $\sim$580 K, the system undergoes ordering of the Fe vacancy and nucleation of the HS Fe$^{2+}$ phase with phase separation at  $\sim$520 K. In going from 520 K to 300 K, the antiferromagnetic order is established and, consequently, the IAD and Fe$^{2+}$ HS fraction gradually increase with cooling [Fig. 5(a)]. Around 300 K, the filamentary and metallic LS Fe$^{2+}$ state starts to grow at the boundary between the neighboring Fe$^{2+}$ HS domains. In going from 300 K to 180 K, the filamentary Fe$^{2+}$ LS fraction gradually increases, likely to be due to chemical pressure from the majority phase, and the Fe$^{2+}$ HS fraction decreases [Fig. 5(b)], which is also consistent with the decrease of the IAD. These behaviors are basically consistent with the previous works on electrical transport showing anomalous temperature dependent resistivity \cite{Guo,Mizuguchi,Ying}. The results are also in agreement with the magnetic \cite{Ming-Hu,Bao2,TexiNMR,SherMuSR}, optical \cite{HomeOpt,YuanOpt} and microscopy measurements \cite{Bendele14}. The most striking result is the increase of the Fe$^{2+}$ HS fraction with cooling from 180 K although the system exhibits a good metallic behavior in the temperature range. This observation suggests that the Fe$^{2+}$ LS state at the interface region again tends to be magnetic but stay metallic as schematically shown in Fig. 5(c). One possible explanation is that the magnetic fluctuation and local spin moment are induced by orbital selective Mott transition in this temperature range. Such a phase has been  proposed recently for K$_x$Fe$_{2-y}$Se$_2$ on the basis of angle-resolved photoemission results \cite{Yi13}. Thus, the apparent increase of the local magnetic moment below 180 K can be explained by the orbital selective localization of the Fe 3$d$ electrons and resulting strong magnetic fluctuations. This assumption is consistent with the recent high pressure measurements underlining the importance of the insulating texture in the superconductivity of A$_x$Fe$_{2-y}$Se$_2$\cite{Zhao14}. Across $T_c$, the interface metallic region with the localized magnetic moment turns into the superconducting LS state, connecting all filamentary metallic phase, as illustrated in Fig. 5(d). It is worth mentioning that the Fig. 5 provides a likely interpretaion of the anomalous behavior of local magnetic moment [Fig. 2] and pre-peak intensity [Fig. 4] as a function of temperature. However, it is difficult to argue that P1 and P2 in Fig. 3 to provide quantitative estimation of the HS and LS phases since the iron vacancy configuration in the majority phase itself evolves with temperature.

\section{Conclusions}\label{S:conclusions}
In summary, we have employed high energy XES and XAS to investigate local magnetic moment and coexisting electronic phases in K$_{x}$Fe$_{2-y}$Se$_2$ as a function of temperature, across the superconducting and phase separation temperatures. We have found coexisting Fe$^{2+}$ electronic phases with the glassy character of K$_{x}$Fe$_{2-y}$Se$_2$ compound. The results provide a clear evidence of the interplay between the electronic and magnetic degrees of freedom for the superconductivity. Across T$_c$, the LS phase increases at the expenses of the interface phase between LS and HS phases.  The HS to LS fraction exchange across T$_c$ is suppressed after annealing despite a substantial presence of the LS configuration, likely to be due to absence of a well defined interface phase. The results suggest that a delicate interplay of local strain between the Mott insulator and metallic filaments with an interface phase is the key for the superconducting phenomena of inhomogeneous K$_{x}$Fe$_{2-y}$Se$_2$ system.

\section{acknowledgments}
We thank A. Sulpice for the magnetic susceptibility measurements on the annealed sample at the end of the thermal cycles. One of us (T.M.) would like to acknowledge hopsitality at the Sapienza University of Rome. This work is partially supported by PRIN2012 (grant numbre 2012X3YFZ2) of MIUR, Italy.


\begin{thebibliography}{40}

\bibitem{Guo}
J. Guo, S. Jin, G.Wang, S.Wang, K. Zhu, T. Zhou, M. He, and X. Chen, 
Phys. Rev. B {\bf 82}, 180520 (2010). 

\bibitem{Mizuguchi}
Y. Mizuguchi, H. Takeya, Y. Kawasaki, T. Ozaki, S. Tsuda, T. Yamaguchi, and Y. Takano,,  Appl. Phys. Lett. {\bf 98}, 042511 (2011).

\bibitem{Ying}
J. J. Ying, X. F. Wang, X. G. Luo, A. F. Wang, M. Zhang, Y. J. Yan, Z. J. Xiang, R. H. Liu, P. Cheng, G. J. Ye, and X. H. Chen, Phys. Rev. B {\bf 83}, 212502 (2011). 

\bibitem{Ming-Hu}
M.H. Fang, H.D. Wang, C.H. Dong, Z.J. Li, C.M. Feng, J. Chen, and H. Q. Yuan, Europhys. Lett. {\bf 94}, 27009 (2011). 

\bibitem{Ryan}
D.H. Ryan, W. N. Rowan-Weetaluktuk, J.M. Cadogan, R. Hu, W.E. Straszheim, S.L. Bud'ko, and P.C. Canfield, Phys. Rev. B {\bf 83}, 104526 (2011).

\bibitem{Bao}
W. Bao, Q.Z. Huang,  G.F. Chen, M. A. Green,  D.M. Wang,  J.B. He, and  Y.-M. Qiu, Chin. Phys. Lett. {\bf 28}, 086104 (2011).

\bibitem{Bao2}
F. Ye, S. Chi, Wei Bao, X. F. Wang, J. J. Ying, X. H. Chen, H. D. Wang, C. H. Dong, and Minghu Fang,  Phys. Rev. Lett.  107, 137003 (2011).

\bibitem{Ricci}
A. Ricci, N. Poccia, G. Campi, B. Joseph, G. Arrighetti, L. Barba, M. Reynolds, M. Burghammer, H. Takeya, Y. Mizuguchi, Y. Takano, M. Colapietro, N. L. Saini, and A. Bianconi, Phys. Rev. B {\bf 84}, 060511(R) (2011). 

\bibitem{Wang2}
Z. Wang, Y. J. Song, H. L. Shi, Z. W. Wang, Z. Chen, H. F. Tian, G. F. Chen, J. G. Guo, H. X. Yang, and J. Q. Li, Phys. Rev B {\bf 83}, 140505(R) (2011).

\bibitem{Chen}
F. Chen, M. Xu, Q. Q. Ge, Y. Zhang, Z. R. Ye, L. X. Yang, Juan Jiang, B. P. Xie, R. C. Che, M. Zhang, A. F. Wang, X. H. Chen, D. W. Shen, X. M. Xie, M. H. Jiang, J. P. Hu, D. L. Feng, Phys. Rev. X {\bf 1}, 021020 (2011).

\bibitem{Guo2}
J. Guo, X. Chen, C. Zhang, J. Guo, X. Chen, Q. Wu, D. Gu, P. Gao, X. Dai, L. Yang, H. Mao, L. Sun, Z. Zhao, Phys. Rev. Lett. {\bf 108}, 197001 (2012).

\bibitem{Wang}
D. M. Wang, J. B. He, T.-L. Xia, and G. F. Chen, Phys. Rev. B {\bf 83}, 132502 (2011).

\bibitem{Zhang}
A. M. Zhang, J. H. Xiao, J. B. He, D. M. Wang, G. F. Chen, and Q. M. Zhang, Phys. Rev. B {\bf 85}, 024518 (2012).

\bibitem{Yan}
Y. J. Yan, M. Zhang, A. F. Wang, J. J. Ying, Z. Y. Li, W. Qin, X. G. Luo, J. Q. Li, Jiangping Hu, X. H. Chen, Sci. Rep. {\bf 2}, 212 (2012).

\bibitem{Han}
F. Han, B. Shen, Z.-Y. Wang, H.-H. Wen, Philos. Mag. {\bf 92}, 2553 (2012).

\bibitem{TexiNMR}Y. Texier, J. Deisenhofer, V. Tsurkan, A. Loidl,  D. S. Inosov, G. Friemel, and J. Bobroff, Phys. Rev. Lett. {\bf 108}, 237002 (2012).

\bibitem{SherMuSR}Z. Shermadini,  H. Luetkens, R. Khasanov, A. Krzton-Maziopa, K. Conder,  E. Pomjakushina, H-H. Klauss, and A. Amato, Phys. Rev. B {\bf 85}, 100501 (2012). 

\bibitem{HomeOpt}C. C. Homes, Z. J. Xu, J. S. Wen, and G. D. Gu, Phys. Rev. B {\bf 86}, 144530 (2012). 

\bibitem{YuanOpt}R. H. Yuan, T. Dong, Y. J. Song, P. Zheng, G. F. Chen, J. P. Hu, J.Q. Li  and N. L. Wang, Sci Rep. 2, 221 (2012).

\bibitem{KsenMoss}V. Ksenofontov,  G. Wortmann, S. A. Medvedev, V. Tsurkan, J. Deisenhofer, A. Loidl, and C. Felser, Phys. Rev. B {\bf 84}, 180508 (2011). 

\bibitem{Shoemaker12}
D. P. Shoemaker, D. Y. Chung, H. Claus, M. C. Francisco, S. Avci, A. Llobet, 
and M. G. Kanatzidis, Phys. Rev. B {\bf 86}, 184511 (2012).

\bibitem{Bendele14}
M. Bendele, A. Barinov, B. Joseph, D. Innocenti, A. Iadecola, H. Takeya, Y. Mizuguchi, T. Takano, T. Noji, T. Hatakeda, Y. Koike, M. Horio, A. Fujimori, D. Ootsuki, T. Mizokawa, and N. L. Saini, Sci. Rep. {\bf 4}, 5592 (2014).

\bibitem{Oiwake13}
M. Oiwake, D. Ootsuki, T. Noji, T. Hatakeda, Y. Koike, M. Horio, A. Fujimori, N. L. Saini, and T. Mizokawa, Phys. Rev. B {\bf 88}, 224517 (2013).

\bibitem{Simonelli12}
L. Simonelli, N. L. Saini, M. Moretti Sala, Y. Mizuguchi, Y. Takano, H. Takeya, T. Mizokawa, and G. Monaco, Phys. Rev. B {\bf 85}, 224510 (2012).

\bibitem{Ponchut}
C. Ponchut, J. M. Rigal, J. Cl\'ement, E. Papillon, A. Homs, and S. Petitdemange, Journal of Instrumentation {\bf 6}, C01069 (2011).

\bibitem{Glatzel}
P. Glatzel and Uwe Bergmann, Coordination Chemistry Reviews {\bf 249}, 65 (2005).

\bibitem{Cowan}
R.D. Cowan, The Theory of Atomic Structure and Spectra, University
of California Press, Berkeley, (1981).

\bibitem{Vanko}
G. Vank\'o, T. Neisius, G. Moln\'ar, F. Renz, S. K\'arp\'ati, A. Shukla, and
F. M. F. de Groot, J. Phys.  Chem.  B {\bf 110}, 11647 (2006).

\bibitem{Gret13} 
H. Gretarsson, S. R. Saha, T. Drye, J. Paglione,  Jungho Kim, D. Casa, T. Gog, W. Wu, S. R. Julian, and Young-June Kim, Phys. Rev. Lett. {\bf 110}, 047003 (2013).

\bibitem{Iadecola12}
A. Iadecola, B. Joseph, L. Simonelli, A. Puri, Y. Mizuguchi, H. Takeya,
Y. Takano, N.L. Saini, J. Phys.: Condens.  Matter {\bf 24}, 115701 (2012).

\bibitem{Simonelli}
L. Simonelli, N. L. Saini, Y. Mizuguchi, Y. Takano, T. Mizokawa,
G. Baldi and G. Monaco, J. Phys. Condens. Matter {\bf 24}, 415501 (2012).

\bibitem{Joseph}
B. Joseph, A. Iadecola, L. Simonelli, Y. Mizuguchi, Y. Takano, T. Mizokawa, and N. L. Saini, J. Phys. Condens. Matter {\bf 22}, 485702 (2010).

\bibitem{HSeLS}
T. E. Westre, P. Kennepohl, J. G. DeWitt, B. Hedman, K. O. Hodgson, and E. I. Solomon, J. Am. Chem. Soc. {\bf 119}, 6297 (1997).

\bibitem{Chen11}
J. M. Chen, S. C. Haw, J. M. Lee, T. L. Chou, S. A. Chen, K. T. Lu, Y. C. Liang,  Y. C. Lee, N. Hiraoka, H. Ishii, K. D. Tsuei, E. Huang, and T. J. Yang, Phys. Rev. B  {\bf 84}, 125117 (2011).

\bibitem{Kumar11}
R. S. Kumar, Y. Zhang, Y. Xiao, J. Baker, A. Cornelius, S. Veeramalai, P. Chow, C. Chen, and Y. Zhao, Appl. Phys. Lett. {\bf 99}, 061913 (2011).

\bibitem{Bendele13a}
M. Bendele, C. Marini, B. Joseph, G. M. Pierantozzi, A. S. Caporale, A. Bianconi, E. Pomjakushina, K. Conder, A. Krzton-Maziopa, T. Irifune, 
T. Shinmei, S. Pascarelli, P. Dore, N. L. Saini, and P. Postorino,
Phys. Rev. B {\bf 88}, 180506 (2013).

\bibitem{Yi13}
M. Yi, D. H. Lu, R. Yu, S. C. Riggs, J.-H. Chu, B. Lv, Z. K. Liu, M. Lu, Y.-T. Cui,
M. Hashimoto, S.-K. Mo, Z. Hussain, C.W. Chu, I. R. Fisher, Q. Si, and Z.-X. Shen,
Phys. Rev. Lett. {\bf 110}, 067003 (2013).

\bibitem{Zhao14}
P. Gao, R. Yu, L. Sun, H. Wang, Z. Wang, Q. Wu, M. Fang, G. Chen, J. Guo, C. Zhang, D. Gu, H. Tian, J. Li, J. Liu, Y. Li, X. Li, S. Jiang, K. Yang, A. Li, Q. Si, and Z.X. Zhao,
Phys. Rev. B {\bf 89}, 094514 (2014).

\end{thebibliography}
\end{document}